# Suppression of DC term in Fresnel digital holography by sequence subtraction of holograms

Jong-Chol Kang, Chol-Su Kim, Song-Jin Im

Department of Physics, **Kim Il Sung** University,

Ryongnam Dong, Taesong District, Pyongyang, D P R Korea

**Abstract**

An experimental method for suppression of DC term in the reconstructed images from Fresnel digital holograms is presented. In this method, two holograms for the same object are captured sequentially and subtracted. Since these two holograms are captured at different moments, they are slightly different from each other for fluctuations of noises. The DC term is suppressed in the image reconstructed from the subtraction hologram, while the two virtual and real images are successfully reconstructed. This method can be potentially used for the improvement of image quality reconstructed from Fresnel digital holograms.

## 1. Introduction

Holography is known as a method recording the amplitude and phase of object wave. In holography, the interference pattern between reference and object beam is recorded on a photosensitive plate, called hologram. In conventional holography, the chemical processes for photographic plates such as developing and fixing are necessary, so it has limitations for practical applications. Recently sensors with high resolution such as CCD and CMOS are used as new hologram recording materials. Holograms are recorded on image sensors and reconstructed with computer digitally. Digital holography are widely used in many scientific fields including particle size measurement and phase retrieve in interferometry and 3-dimensional shape measurement [1-4]. However, the image quality reconstructed from digital hologram is not good compared to one from conventional photographic plate because the resolutions of CCD and CMOS are not yet so high.

In digital holography, the pixel size of CCD and CMOS should be less than the spatial period of holographic pattern, which depends on the angle between reference and object beam. The angle is normally limited to within a few of degree. In in-line hologram (Gabor hologram), where reference and object beam are almost parallel, the pixel resolution of image detectors is high enough to distinguish the holographic pattern. However, zero-order term, real and virtual images are overlapped in reconstructed image. So there must be methods to separate these three terms clearly. In off-axis digital hologram zero-order term, real and virtual images can be separated but the separation between them is small because of the limitation of CCD pixel resolution. Furthermore, they are sometimes overlapped partially, so some useful information in the reconstructed image could be lost.

There are mainly two methods to suppress DC term; experimental and image processing method. The most well-known experimental method is phase shifting digital holography proposed by Takaki[10] and Yamaguchi[9,11], where several holograms are captured while changing the phase differences between reference and object waves. Using this method, twin images are separated exactly and the DC term is highly suppressed.

However, this method needs to use phase-shifting devices like piezo-activated mirror driven by up-stepping or down-stepping voltages. So its practical applications are limited in studying dynamic phenomena. Kreis and Jüptner proposed a good image processing method to suppress the DC term [5, 6, 12]. In this method, the recorded hologram multiplied with the reference wave is averaged and then the average intensity is subtracted from each stored hologram intensity value, yielding the modified digital hologram. In the image reconstructed from this modified hologram, the DC term is successfully removed. Here several steps for image processing are needed such as multiplying, averaging and subtraction. Beyond these there have been several image processing methods based on filtering of Fourier spectrum[8].

In this paper, a new experimental method for suppression of DC term is presented. In this method, two holograms for the same object are captured sequentially and subtracted. The DC component is suppressed in the image reconstructed from the subtraction hologram, while the two virtual and real images are successfully reconstructed. A similar method to this one was already proposed by Demoli, Mestrovic and Sovic[7], Subtraction digital holography. Here, a stochastic phase is introduced and subtraction of the digital hologram recorded with this additional phase leads to a suppression of the unwanted terms. However this method requires the modification of the holographic arrangement with shutter and phase modulator. The following method need only two holograms captured sequentially without any manipulation of holographic arrangement and Fourier spectrum. This method is simple and comfortable to remove the unwanted DC term and to improve the qualities of the reconstructed images.

We first demonstrate our experimental result and then explain it with computer simulation.

## 2. Basic principle of digital holography

The intensity distribution $I_H(x, y)$ of hologram formed by interference between reference wave $A(x, y)$ and object wave $a(x, y)$ can be written by

$$I_H(x, y) = (A+a) \cdot (A+a)^* = |A|^2 + |a|^2 + A^*a + Aa^*, \qquad (1)$$

where $A^*$ and $a^*$ are the complex conjugates of the reference and object waves, respectively. The sum of the first two terms of this equation is DC term (or zero-order term) and slowly varying in time or space, not depending on the phase of $A$ and $a$. The last two terms are interference terms. They are sensitive to the phase difference between $A$ and $a$ and will be processed to retrieve the phase information.

Reconstructing the object wave can be achieved by illuminating the hologram with the original reference wave. It is expressed mathematically by

$$A \cdot I_H = A \cdot (A+a) \cdot (A+a)^* = A \cdot (|A|^2 + |a|^2) + A^2 a + A^2 a^* \qquad (2)$$

The first term on the right part of Eq. (2) corresponds to the reconstruction of the zero order term and should be eliminated as completely as possible. The second term corresponds to the virtual image and restores the object wave. The third term corresponds to a wave outgoing from the hologram and converging to the real image of the object, which appears as a mirror image of the virtual one.

In the upper equations, $a$ is not directly the object wave, but the diffracted wave propagated over certain distance from object. Therefore, in digital holography, it must be considered the Fresnel diffraction in order to calculate the amplitude and the phase of object wave. It can be described with Fresnel transformation in recording and with inverse Fresnel transformation in reconstruction.

In digital Holography, the hologram is captured and digitized by frame-grabber. In digitization, the Fresnel diffraction $a_{\text{CCD}}$ on CCD from the object can be expressed as follows;

$$a_{\text{CCD}}(m\Delta\xi, n\Delta\eta) = -i\exp(2i\pi d/\lambda) \cdot DF_{Fresnel}T[a_{\text{obj}}(k\Delta x, l\Delta y)] \qquad (3)$$

where $\Delta\xi, \Delta\eta$ are the sampling intervals in CCD plane and $\Delta x, \Delta y$ the sampling intervals in object. $m, n, k, l$ are integers and $d$ is the distance between CCD and object. $a_{\text{obj}}$ is the source wave of the object. The discrete Fresnel transform $DF_{Fresnel}T$ in this equation takes the form

$$\begin{aligned} DF_{Fresnel}T[a_{\text{obj}}(k\Delta x, l\Delta y)] &= \frac{1}{\sigma^2}\exp\{\frac{i\pi}{\sigma^2}[m^2\Delta\xi^2 + n^2\Delta\eta^2]\} \cdot \\ &\cdot DFT[a_{\text{obj}}(k\Delta x, l\Delta y) \cdot \exp\{\frac{i\pi}{\sigma^2}[k^2\Delta x^2 + l^2\Delta y^2]\}]_{m,n} \end{aligned} \qquad (4)$$

where $\sigma = \sqrt{\lambda d}$ and $DFT$ is the Discrete Fourier Transform. The wave $a_{\text{CCD}}$ interferes with the reference wave on CCD and forms the hologram $I_H$.

In the case of a plan wave illumination, the wave $a'_{\text{obj}}$ diffracted from the hologram can be expressed by the inverse discrete Fresnel transform.

$$a'_{\text{obj}}(k\Delta x, l\Delta y) = i\exp(-2i\pi d'/\lambda) \cdot DF_{Fesnel}T^{-1}[I_H(m\Delta\xi, n\Delta\eta)] \qquad (5)$$

$$\begin{aligned} DF_{Fresnel}T^{-1}[I_H(m\Delta\xi, n\Delta\eta)] &= \frac{1}{\sigma^2}\exp\{-\frac{i\pi}{\sigma^2}[k^2\Delta x^2 + l^2\Delta y^2]\} \cdot \\ &\cdot DFT^{-1}[I_H(m\Delta\xi, n\Delta\eta) \cdot \exp\{\frac{i\pi}{\sigma^2}[m^2\Delta\xi^2 + m^2\Delta\eta^2]\}]_{k,l} \end{aligned} \qquad (6)$$

where $DFT^{-1}$ is the inverse Discrete Fourier Transform and $d'$ is the reconstruction distance between hologram and screen. If the reconstruction distance is equal to the distance between CCD and object ($d = d'$) during recording, the most clear real image is obtained in the reconstructed intensity of $a'_{\text{obj}}$.

$DFT$ or $DFT^{-1}$ can be easily and efficiently calculated with Fast Fourier Transformation (FFT) algorithm.

### 3. Suppression of DC term by sequence subtraction of holograms

#### 3.1. Experimental setup

Figure 1 shows the experimental setup for recording the digital Fresnel hologram. A double frequency solid state laser (532nm, 50mW) has been used as the light source. The laser beam is split into two equal halves by a nonpolarizing beam splitter cube(BS). Digital holograms are recorded with a monochrome CCD sensor (1024×1024 pixel, 6.7μm×6.7μm) and the test specimen were dice with dimension of 15mm. The reference beam is collimated to illuminate the CCD with parallel beam and the object beam is expanded by a convex lens to illuminate the object. The light diffracted by the object interferes with the reference beam on the CCD. The neutral-density filter (NF) is used to control the intensity ratio between reference and object beam. Plane mirrors ($M_1$, $M_2$, $M_3$) mounted onto adjustable precision kinematic mounts are used to adjust the direction of the beams. The interference pattern formed by the reference beam and the beam diffracted by the object is captured by the CCD sensor and digitized by frame grabber card. The angle separation between the reference and object beam is up to 4° and the distance from the object to the CCD is 90cm~95cm.

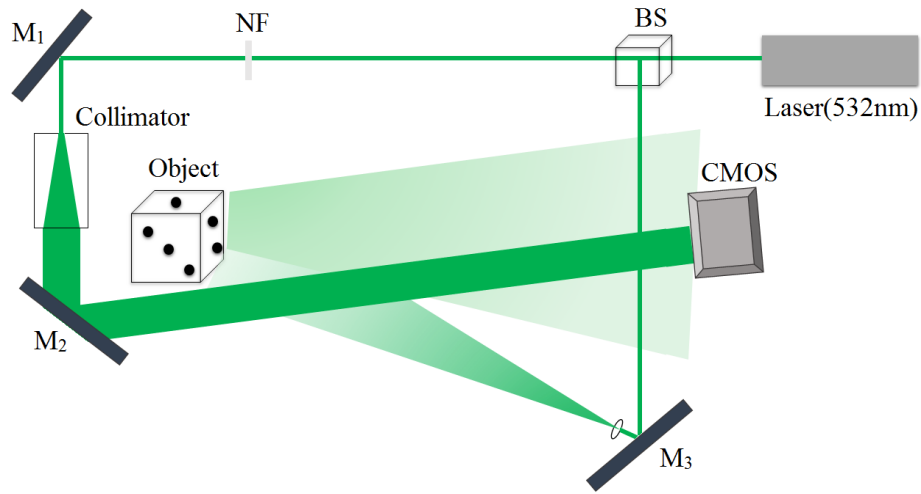

Fig.1. the optical setup for recording Fresnel hologram.

### 3.2. Reconstruction from normal Fresnel hologram

Figure 2 shows a digital Fresnel hologram recorded by the optical configuration shown in Figure 1 and the intensity of the object wave reconstructed from the hologram. In the intensity field of the reconstructions we recognize a bright central square, namely, the DC term of the Fresnel transform, which is much brighter than the real or virtual images. In digital Fresnel hologram the separation between DC term and real or virtual image is small because of the limitation of CCD pixel resolution. Furthermore, they are sometimes overlapped partially, so some useful information in the reconstructed image could be lost.

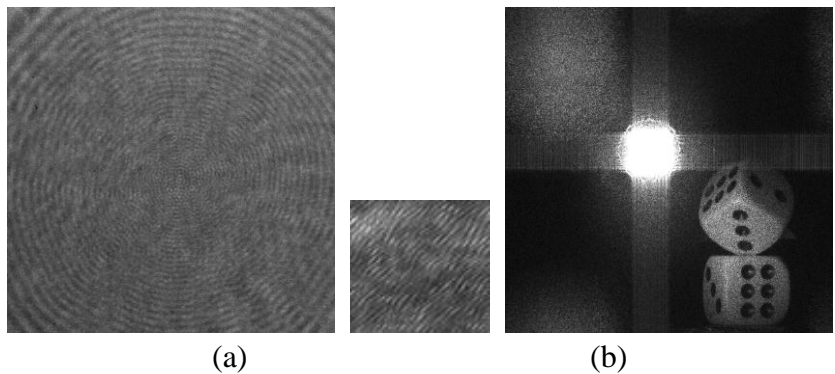

(a)                              (b)

Fig. 2. (a) Digital Fresnel hologram, (b) a magnification of a part of (a), (c) reconstructed intensity of dice

### 3.3. Reconstruction from sequence subtraction of holograms

If two images for the same scene are digitally captured at different moments, they are slightly different from each other for fluctuations of noises in electronics. These fluctuations result in stochastic fluctuations of phase in the case of holographic recording.

Sequence subtraction of two holograms can be realized with the following example codes in Image Acquisition Toolbox in MATLAB.

```
IH1 = double (getsnapshot(vid));
IH2 = double (getsnapshot(vid));
IH  = IH2-IH1;
vid = videoinput ('winvideo', 1, 'RGB16_1024x1024');
```

The recording time is doubled. But the entire time is tolerable because there is no need for any additional hardware manipulation. The resultant subtraction hologram seems to be zero because it was obtained from subtraction between two equal images, "HI1" and "HI2", for the same static scene, and therefore its Fresnel transformation seems to be meaningless. But they are recorded at different moments within dozens of milliseconds, so they are slightly different from each other because of fluctuation of noises.

Figure 3(a) and 3(b) show the subtraction IH between two holograms IH1 and IH2 captured sequentially and its magnification. As shown in the figure, the holographic interference pattern remain even though it is obtained from subtraction between equal holograms and the subtraction hologram has uniform background intensity, comparing it with Figure 2. Figure 3(c) shows the intensity reconstruction from subtraction hologram. The DC term is successfully suppressed in the reconstruction image with the proposed approach.

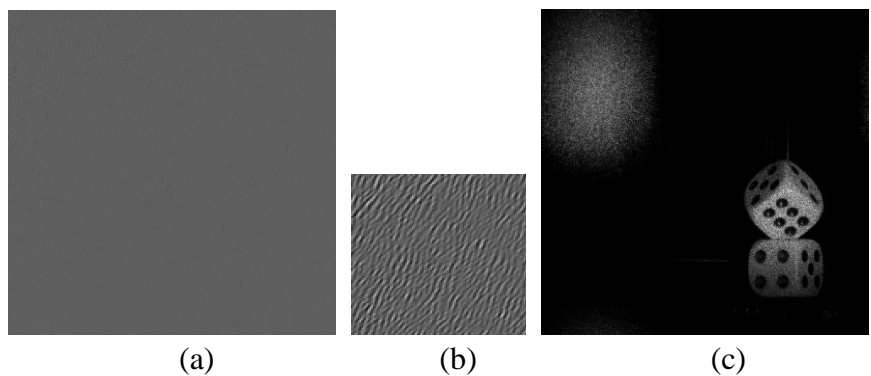

(a)　　　　　　　(b)　　　　　　　(c)

Fig. 3. (a) subtraction hologram, (b)magnification of (a), (c) reconstructed intensity

### 4. Explanation on experimental results with computer simulation

The suppression of the DC term in the reconstructed intensity can be demonstrated with computer simulation of Fresnel digital hologram. The simulation is proceeded in MATLAB.

If an object with diffuse surface is illuminated by collimated light wave or by even random light wave, the light wave is scattered into all the possible directions. This scattering character of the reflected (or transmitted) light wave can be simulated with random phase mask.

First the image to be reconstructed is padded onto a zero-image with certain column and row pixel number. (see Figure 4).

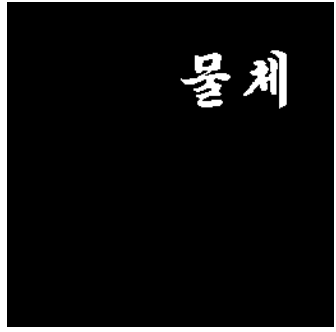
Fig. 4 Zero padding image

The modulated light wave $a_{obj}(x,y)$ scattered from a digital image (the padded image) $a(x,y)$ can be described as multiplication with a random phase mask $\phi(x,y)$.

$$a_{obj}(x,y) = a(x,y)\exp[i\phi(x,y)] \qquad (7)$$

The Fresnel transformation $a_{CCD}(\xi,\eta)$ of $a_{obj}(x,y)$ interferes with a planar reference wave $A(\xi,\eta)$ on CCD and forms hologram $I_H(\xi,\eta)$.

$$A(\xi,\eta) = A_0 \exp[i2\pi(p\xi+q\eta)] \qquad (8)$$

$$\begin{aligned}I_H(\xi,\eta) &= |a_{CCD}(\xi,\eta)+A(\xi,\eta)|^2 \\ &= |a_{CCD}(\xi,\eta)|^2 + |A(\xi,\eta)|^2 \\ &\quad + a_{CCD}^*(\xi,\eta)A(\xi,\eta) + a_{CCD}(\xi,\eta)A^*(\xi,\eta)\end{aligned} \qquad (9)$$

The two first terms in right side of Eq. (9) make the DC term in reconstruction. If the hologram obtained with Eq. (9) is Fresnel-transformed inversely, the real and virtual images are reconstructed, which are symmetrical to the DC term and conjugated to each other. The positions of the reconstructed mark images can be controlled by selecting the position of mark image in zero padding and the angular spectrums $\xi,\eta$ of the planar reference wave during recording process. Figure 5 shows a simulated Fresnel hologram and the reconstructed intensity.

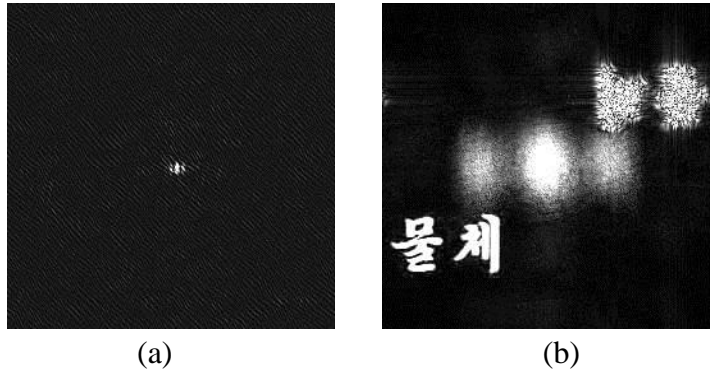

(a)          (b)
Fig. 5 (a) Simulated Fresnel hologram (256×256 pixel), (b) reconstructed intensity

Subtraction hologram is realized by difference between two holograms with different random phase masks.

Figure 6 shows two different holograms (a, b) with different phase masks, the subtraction hologram(c) from (a) and (b) and the reconstructed image from (c).

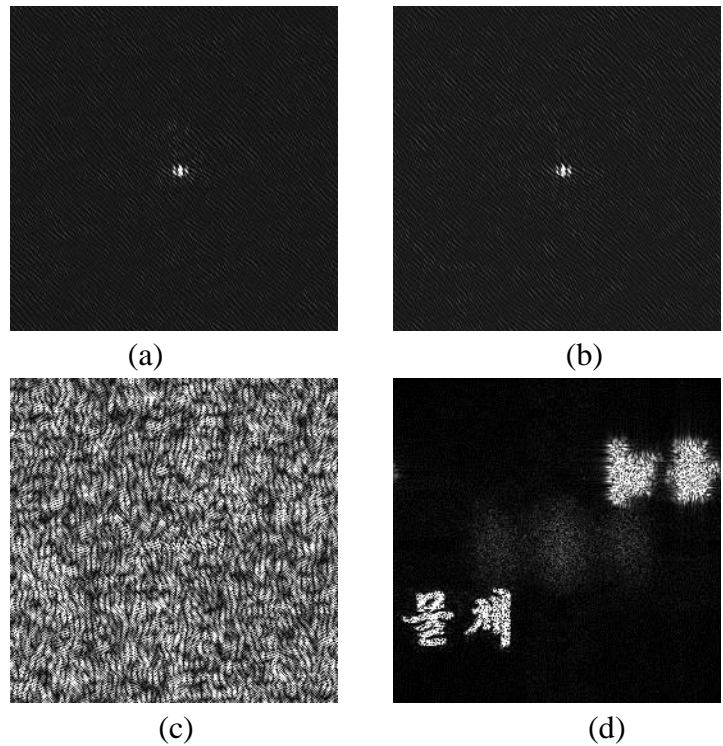

Fig.6 (a), (b) simulated holograms with different phase masks, (c) subtraction hologram, (d) reconstructed intensity

Comparing Figure 5 (b) with Figure 6 (d), it is clear that the DC term in the reconstructed intensity from subtraction hologram is successfully suppressed. The difference in phase noises do not affect the image quality in the case of interference, while the background components are removed.

## 5. Application to digital holographic interferometry

The proposed approach can be applied to double-exposure and time-averaged digital holographic interferometry.

We demonstrate the possibility for application of sequence subtraction of holograms to double exposure digital holographic interferometry.

The test specimen was a circular brass plate (0.3 mm thickness, 50mm diameter) held on the optical table by clamping the plate on edge. The center of the plate was activated on the backside by a piezo element driven by changing voltages. It would induce a deformation in the plate.

The two holograms from which the object waves are reconstructed are captured by the following MATLAB codes.

```
    vid = videoinput ('winvideo', 1, 'RGB16_1024x1024');
    IH1= double(getsnapshot(vid));
    IH2 = double(getsnapshot(vid));
    IH_B = IH2-IH1;
Activation of test object for deformation
    IH3 = double(getsnapshot(vid));
    IH4 = double(getsnapshot(vid));
    IH_A = IH4-IH3;
```

The interferogram is obtained by comparison between two object waves reconstructed from IH_B and IH_A before and after deformation. Figure 7 shows a double-exposure holographic interferogram (a, b) without DC term and the unwrapped phase map of it(c).

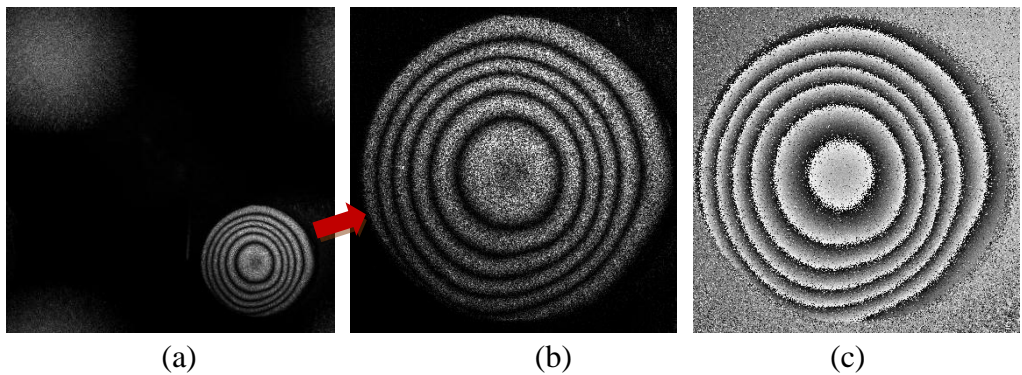

(a)    (b)    (c)

Fig. 7 (a, b) digital holographic interferogram, (c) unwrapped phase map

Figure 7 shows that DC term is successfully suppressed by subtraction of holograms captured sequentially and the reconstructed image is not degraded in this process, not only in intensity, but also in phase map.

### 6. Conclusions

This paper presents an approach, namely, the sequence subtraction of hologram, for suppression of DC term in digital Fresnel hologram. Experimental results demonstrate that the proposed method is simple and useful for suppression of DC term in digital Fresnel hologram. Since two holograms have to be captured for subtraction the recording time is doubled. But the entire time is tolerable because there is no need for any additional hardware manipulation.

The suppression of the DC term in the reconstructed intensity can also be demonstrated with computer simulation of Fresnel digital hologram. The object wave scattered from diffuse surface is simulated by multiplication with a random phase mask. Simulation results show that the DC term in the reconstructed intensity from subtraction hologram is successfully suppressed. The difference in phase noises do not affect the image quality in the case of interference, while the background components are removed.

The proposed approach can be applied to observation for deformation and vibration by double-exposure and time-averaged digital holographic interferometry. The reconstructed image is not degraded in subtraction hologram, not only in intensity, but also in phase map.

### References
1. U. Schnars, W. Jüpner, "Digital Holography", Spriner-Verlag, 2005
2. W. Osten, A. Faridian, P. Gao, K. Körner, D. Naik, G. Pedrini, A. K. Singh, M. Takeda, and Marc Wilke1, "Recent advances in digital holography [Invited]" Applied Optics, Vol. 53, No. 27, 2014, G44-G63
3. H. J. Tiziani and G. Pedrini, "From speckle pattern photography to digital holographic interferometry [Invited]" Applied Optics, Vol. 52, No. 1, 2013, 30-44
4. T. Kreis, "Handbook of Holographic Interferometry", Wiley-VCH Verlag, 2005

5. U. Schnars, W. Jüpner, "Digital Holography", Spriner-Verlag, 2005, 56-61

6. T. Kreis, "Handbook of Holographic Interferometry", Wiley-VCH Verlag, 2005, 105-107

7. N. Demoli, J. Mestrovic and I. Sovic, "Subtraction digital holography" Applied Optics, 42:798–804, 2003.

8. C. Depeursinge, "Digital holography applied to microscopy" in Digital holography and three-dimensional display, T.-C. Poon, ed. (Springer-Verlag, 2006), Chap. 4, 95-143

9. I. Yamaguchi, "Phase-shifting digital holography" in Digital holography and three-dimensional display, T.-C. Poon, ed. (Springer-Verlag, 2006), Chap. 5, 145-171

10. Takaki Y, Kawai H, Ohzu H. Hybrid holographic microscopy free of conjugate and zero-order images. Appl Opt 1999;38:4990–6

11. Yamaguchi I, Zhang T. Phase-shifting digital holography. Opt Lett 1998;23:1221–3

12. Kreis T, Jüptner W, "Suppression of the DC term in digital holography", Opt Eng 1997;36:2357–60